\documentclass[12pt]{article}
\usepackage{cite,epsfig,amsfonts,amsmath}
\usepackage{slashed,amssymb,graphicx,color}
\begin{document}

\baselineskip=.22in
\renewcommand{\baselinestretch}{1.2}
\renewcommand{\theequation}{\thesection.\arabic{equation}}
~\vspace{12mm}
%\begin{flushright}
%{\tt arXiv:YYMM.NNNN}
%\end{flushright}

%%%%%%%%%%%%%%%%%%%%%%%%%%%%%%%%%%%%%%%%%%%
\newcommand{\be}{\begin{equation}}
\newcommand{\ee}{\end{equation}}
\newcommand{\bear}{\begin{eqnarray}}
\newcommand{\eear}{\end{eqnarray}}
\newcommand{\beal}{\begin{align}}
\newcommand{\eeal}{\end{align}}
\newcommand{\ba}{\begin{array}}
\newcommand{\ea}{\end{array}}
\newcommand{\nn}{\nonumber}
\newcommand{\tcb}{\textcolor{blue}}
\newcommand{\tcr}{\textcolor{red}}
\newcommand{\tcg}{\textcolor{green}}
\def\bra#1{\langle #1 |}
\def\ket#1{|#1 \rangle}
%%%%%%%%%%%%%%%%%%%%%%%%%%%%%%%%%%%%%%%%%%%

\begin{center}
{{\Large \bf Notes on Supersymmetry Enhancement of ABJM Theory }
}\\[20mm]
{{O-Kab Kwon$^1$, ~~Phillial Oh, ~~Jongsu Sohn}\\[3mm]
{\it Department of Physics, $^1$BK21 Physics Research Division,
and Institute of Basic Science\\
Sungkyunkwan University, Suwon 440-746, Korea}\\
{\tt okab@skku.edu, ploh@skku.edu, jongsusohn@skku.edu} }
\end{center}
\vspace{10mm}

\begin{abstract}
We study the supersymmetry enhancement of ABJM theory. Starting from
a ${\cal N}=2$ supersymmetric Chern-Simons matter theory with gauge
group U(2)$\times$U(2) which is a truncated version of the ABJM
theory, we find by using the monopole operator that there is
additional ${\cal N}=2$ supersymmetry related to the gauge group.
We show this additional supersymmetry can combine with ${\cal
N}=6$ supersymmetry of the original ABJM theory to an enhanced
 ${\cal N}=8$ SUSY  with gauge group U(2)$\times$U(2)
in the case $k=1,2$. We also discuss the supersymmetry enhancement
of the ABJM theory with U($N$)$\times$U($N$) gauge group and
find a condition which should be satisfied by the monopole operator.
\end{abstract}

%{\it{Keywords}} :

\newpage

\setcounter{equation}{0}
\section{Introduction}\label{section1}

There has been remarkable recent progress in understanding the
worldvolume theory of coincident M2-branes. This was initiated by
Bagger and Lambert~\cite{Bagger:2006sk} and
Gustavsson~\cite{Gustavsson:2007vu} (BLG) who found an ${\cal N}=8$
Chern-Simons matter theory based on 3-algebra. Under the assumption
for Euclidean metric in the 3-algebra, the gauge group of the
BLG theory is restricted to SO(4). So the BLG theory can be
reformulated as an ordinary Chern-Simons gauge theory with
SU(2)$\times$SU(2) gauge group having opposite Chern-Simons levels
$k$ and $-k$~\cite{VanRaamsdonk:2008ft}. Inspired by BLG theory and
subsequent developments, Aharony, Bergman, Jafferis, and Maldacena
(ABJM) proposed ${\cal N}=6$ Chern-Simons matter theory with
U($N$)$\times$U($N$) gauge group~\cite{Aharony:2008ug}. The ABJM
theory is believed as a low energy effective theory of multiple
M2-branes on orbifold ${\bf C}^4/{\bf Z}_k$. According to the
developments related to the M2-brane effective actions, the
Chern-Simons matter theories with higher number (${\cal N}\ge 4$) of
supersymmetry were also
constructed~\cite{Gaiotto:2008sd,Hosomichi:2008jd,Hosomichi:2008jb}.

The ABJM theory with gauge group SU(2)$\times$SU(2) is equivalent to
the BLG theory as proved in Ref.~\cite{Aharony:2008ug}.
So it has ${\cal N}=8$ supersymmetry regardless the Chern-Simons level $k$.
Unlike the SU(2)$\times$SU(2) case, the ABJM theory has ${\cal N}=6$
supersymmetry for generic $k$. It was conjectured, however, that
the ABJM theory has the additional ${\cal N}=2$ supersymmetry and becomes
${\cal N}=8$ theory at $k=1,2$~\cite{Aharony:2008ug}.

The purpose of this paper is to find the additional ${\cal N}=2$
supersymmetry explicitly and prove the conjecture for supersymmetry
enhancement in ABJM theory with U(1)$\times$U(1) and U(2)$\times$U(2)
gauge groups. We also propose a general formulation for the additional
supersymmetry in ABJM theory with U($N$)$\times$U($N$) gauge group.
To do so, we introduce a local operator $T^{ab}_{\hat a\hat b}$ (or
$T^{\hat a\hat b}_{ab}$) in the supersymmetry transformation rules,
where $a,b$ and $\hat a,\hat b$ are the
gauge indices of ${\rm U}(N)_L$ and ${\rm U}(N)_R$ gauge groups
respectively. After some calculations we determine the condition for
$T$, which gives the additional ${\cal N}=2$ supersymmetry. Since
there are two gauge groups in ABJM theory, the matter fields are in
bifundamental or anti-bifundamental representations, which are
interchanged with each other with the action of $T$ on these fields.
For instance, a bifundamental scalar $Y^A$ is changed to
an anti-bifundamental scalar $TY^A$ due to the index structure of $T$.
Actually $T$ corresponds to the monopole operator (often called 't
Hooft operator), which was suggested in Ref.~\cite{Klebanov:2008vq}.
For an explicit study of monopole operators in the ABJM theory and
related topics, see~\cite{Berenstein:2008dc,Hosomichi:2008ip,
Park:2008bk,Imamura:2009ur, Kim:2009wb,SheikhJabbari:2009kr,
Berenstein:2009sa, Benna:2009xd}

It is interesting that the supersymmetry parameter for the
additional ${\cal N}=2$ supersymmetry includes gauge indices and
crucially depends on the gauge group of the theory. In this sense,
the additional supersymmetry in ABJM theory is an exceptional one in
supersymmetric gauge theories.

For U(1)$\times$U(1) case, $T$ becomes the abelian monopole operator
as we will see in the subsection \ref{U1U1section}, and the
additional supersymmetry is allowed for $k=1,2$ cases due to the
orbifold structure of the transverse space. On the other hand, for
SU(2)$\times$SU(2) case, $T$ is expressed as the product of the
SU(2) invariant tensors $\epsilon^{ab}$ and $\epsilon_{\hat a\hat
b}$, which are independent of the worldvolume coordinates, and the
additional supersymmetry always exist for any value of $k$.
Therefore the additional supersymmetry seems to be allowed only for
$k=1,2$ cases in U($N$)$\times$U($N$)  or SU($N$)$\times$SU($N$) ($N\ge 3$)
gauge groups, which are composed of U(1) and SU(2)
parts.

The rest of this paper is organized as follows. In section 2, we
introduce a superconformal Chern-Simons matter theory which is a
truncated version of the ABJM theory but has the same supersymmetry
enhancement properties  with minimal number of matter fields. We
call this theory as the {\it minimal model}. The model has the same
forms of the kinetic terms for scalars and fermions and the
Chern-Simons terms. And the matter field part is composed of two
complex scalars and fermions and so the fermionic and bosonic
potentials are different from those of ABJM theory. We explicitly
show ${\cal N}=2$ supersymmetry of the model having $U(1)_R$
symmetry and find the additional ${\cal N}=2$ supersymmetry for
U(1)$\times$U(1), SU(2)$\times$SU(2), and U(2)$\times$U(2) cases. In
Appendix A, we verify the supersymmetric invariance of the
Lagrangian of the minimal model. In section 3, we prove the
conjecture for the supersymmetry enhancement in ABJM theory for
U(1)$\times$U(1) and U(2)$\times$U(2) cases at $k=1,2$, and suggest
a possible supersymmetry transformation rules for the additional
${\cal N}=2$ supersymmetry and corresponding condition in $T$ for
the general U($N$)$\times$U($N$) or SU($N$)$\times$SU($N$) cases. In
Appendix B, we show that the procedure for the minimal model can
also be applied to ABJM theory. We conclude in section 4 with brief
summary and discussion.
\\

\noindent
{\bf Note Added:} While this paper was being completed,
 a paper arXiv:0906.3568 [hep-th]\cite{Gustavsson:2009pm} appeared,
which also deals with supersymmetry enhancement of ABJM theory
with general gauge group based on 3-algebra.

\section{Supersymmetry Enhancement of a Minimal Model}

Before taking into account the supersymmetry enhancement of ABJM theory,
we consider supersymmetry enhancement of a minimal model, which
is a ${\cal N}=2$ superconformal Chern-Simon matter theory and
has the same supersymmetry enhancement behaviors with those of ABJM theory.
The model has the same kinetic terms
for scalars and fermions, Chern-Simons terms with gauge group
U($N$)$\times$U($N$)(or SU($N$)$\times$SU($N$)) with the ABJM theory.
However, the minimal model has two complex scalars and fermions
with global SU(2)$\times$U(1) symmetry, while ABJM theory has
four complex scalars and fermions with global SU(4)$\times$U(1) symmetry.
So the fermionic and bosonic potentials
of this ${\cal N}=2$ Chern-Simons theory are different from those of
${\cal N}=6$ ABJM theory.

Fields in the ${\cal N}=2$ minimal model are composed of two gauge
fields $A_\mu$ and $\hat A_\mu$, two bifundamental  bosonic fields
$Z^A$ and fermionic fields $\psi_A$ with $A=1,2$, and their
Hermitian conjugates $Z_A^\dagger$ and $\psi^{\dagger A}$
respectively. $Z^A$ and $\psi^{\dagger A}$ with upper indices
($Z_A^\dagger$ and $\psi_A$ with lower indices) are in the ${\bf 2}$
(${\bf {\bar 2}}$) representation of the global SU(2). The gauge and
matter fields have gauge group indices for U($N$)$\times$U($N$) (or
SU($N$)$\times$SU($N$)) as $A^a_{~b}$, ${\hat A}^{\hat a}_{~\hat
b}$, $Z^{a}_{~\hat b}$, and $\psi^a_{~\hat b}$. And the conjugate
fields are represented as $Z^{\dagger\hat a}_{~~b}$ and
$\psi^{\dagger\hat a}_{~~b}$. Then the action with global
SU(2)$\times$U(1) symmetry is given by\footnote{We choose
(2+1)-dimensional gamma matrices which satisfy
$\gamma^\mu\gamma^\nu=\eta^{\mu\nu} +
\epsilon^{\mu\nu\rho}\gamma_\rho$ as $ \gamma^0= i\sigma^2,
\gamma^1=\sigma^1$, and $\gamma^2=\sigma^3$. The suppressed spinor
indices are expressed by $\xi \chi\equiv \xi^\alpha\chi_\alpha$ and
$\xi\gamma^\mu\chi=\xi^\alpha\gamma_{\alpha}^{\mu\,\beta}\chi_\beta$
for the two component spinors $\xi$ and $\chi$. The conventions of
gauge indices for bosonic and fermionic fields are same as those in
Ref.~\cite{Benna:2008zy}.}
\begin{align}\label{dact}
S =\int d^3x\,\left({\cal L}_0 + {\cal L}_{{\rm CS}} -V_{{\rm ferm}}
-V_{{\rm bos}} \right)
\end{align}
with
\begin{align}
{\cal L}_0 &= {\rm tr}\left(-D_\mu Z_A^\dagger D^\mu Z^A +
i\psi^{\dagger A} \gamma^\mu D_\mu \psi_A\right),
\label{dakin} \\
{\cal L}_{{\rm CS}} &= \frac{k}{4\pi}\,\epsilon^{\mu\nu\rho}\,{\rm tr}
\left(A_\mu \partial_\nu A_\rho +\frac{2i}{3}A_\mu A_\nu A_\rho
- \hat{A}_\mu \partial_\nu \hat{A}_\rho
-\frac{2i}{3}\hat{A}_\mu \hat{A}_\nu \hat{A}_\rho \right),
\label{dacs} \\
V_{{\rm ferm}} &= \frac{2\pi i}k{\rm tr}\Big( Z_A^\dagger Z^A\psi^{\dagger
B}\psi_B -Z^A Z_A^\dagger\psi_B \psi^{\dagger B}
+2Z^AZ_B^\dagger\psi_A\psi^{\dagger B} -2Z_A^\dagger
Z^B\psi^{\dagger A}\psi_B\Big),
\label{fpot} \\
V_{{\rm bos}} &=\frac{4\pi^2}{k^2}{\rm tr}\Big(
Z^\dagger_AZ^AZ^\dagger_BZ^BZ^\dagger_CZ^C
+Z^AZ^\dagger_AZ^BZ^\dagger_BZ^CZ^\dagger_C
-2Z^AZ^\dagger_BZ^BZ^\dagger_AZ^CZ^\dagger_C \Big).
\label{bpot}
\end{align}
where the covariant derivatives are defined as
\begin{align}\label{covD}
D_\mu Z^A &=\partial_\mu Z^A +iA_\mu Z^A -iZ^A \hat{A}_\mu,
\nn \\
D_\mu Z^\dagger_A &=\partial_\mu Z^\dagger_A +i{\hat A}_\mu Z^\dagger_A
-iZ^\dagger_A A_\mu.
\end{align}
We can also obtain the action (\ref{dact}) by turning off two scalars and
two fermions in the ${\cal N}=2$ superspace formalism for BLG theory given
in Ref.~\cite{Benna:2008zy}. The F-term potentials vanish when we consider
two complex fields only.

The action (\ref{dact}) is invariant under ${\cal N}=2$ supersymmetry
transformation,
\begin{align}\label{dN2susy}
\delta Z^A&= i\varepsilon^\dagger\epsilon^{AB}\psi_B,
\nn \\
\delta Z^\dagger_A&= i\epsilon_{AB}\psi^{\dagger B}\varepsilon,
\nn \\
\delta\psi_A&=\epsilon_{AB}D_\mu Z^B\gamma^\mu\varepsilon +
\epsilon_{AB} N^B\varepsilon,
\nn \\
\delta\psi^{\dagger A}&=-\varepsilon^\dagger\epsilon^{AB}
\gamma^\mu D_\mu Z^\dagger_B
+ \varepsilon^\dagger\epsilon^{AB} N_{B}^\dagger,
\nn \\
\delta A_\mu&=-\frac{2\pi}{k}\left(\varepsilon^\dagger\epsilon^{AB}
\gamma_\mu
\psi_BZ^\dagger_A + \epsilon_{AB}Z^A\psi^{\dagger B}\gamma_\mu\varepsilon
\right),
\nn \\
\delta \hat A_\mu&=-\frac{2\pi}{k}\left(\varepsilon^\dagger
\epsilon^{AB}Z^\dagger_A
\gamma_\mu\psi_B + \epsilon_{AB}\psi^{\dagger B}\gamma_\mu Z^A\varepsilon
\right),
\end{align}
where we define
\begin{align}\label{dfnN}
N^A\equiv\frac{2\pi}{k}\,
\left(Z^B Z^\dagger_BZ^A-Z^AZ^\dagger_BZ^B\right),
\end{align}
and $\epsilon^{AB}$ and $\epsilon_{AB}$ are the invariant tensors
of the global SU(2) symmetry with $\epsilon^{12}=\epsilon_{12}=1$.
$\varepsilon$ and  $\varepsilon^\dagger$ are the complex spinor parameter and its
complex conjugate respectively. We prove the supersymmetry transformation
rules (\ref{dN2susy}) in Appendix \ref{N2global}.

The action (\ref{dact}) has the additional ${\cal N}=2$
supersymmetry depending on gauge group. As we will see later, the
supersymmetry enhancement behaviors of the ${\cal N}=2$ minimal model are
exactly same with those of ABJM theory. We find the additional
supersymmetry for the model (\ref{dact}) with gauge groups,
U(1)$\times$U(1), SU(2)$\times$SU(2), and U(2)$\times$U(2) cases.

\subsection{U(1)$\times$U(1) case}\label{U1U1section}

In U(1)$\times$U(1) case, we combine the two gauge fields
$A_\mu$ and ${\hat A}_\mu$ into
\begin{align}\label{Apm}
A_{\mu}^{\pm}\equiv A_\mu \pm \hat A_\mu,
\end{align}
where $A_\mu^{+}$ does not interact with all matter fields and the
corresponding flux is quantized via Chern-Simons terms, and
$A_\mu^-$ is the ${\rm U}(1)_{A^-}$ (from now on, we denote it as
${\rm U}(1)_{A^-}$) gauge field which will be used inside the
covariant derivatives (\ref{covD}). The matter fields
$(Z^A,\psi^{\dagger A})$ in ${\bf 2}$ representation of the global
SU(2) have the ${\rm U}(1)_{A^-}$ charges ($+,-$), while their
Hermitian conjugates $(Z^\dagger_A,\psi_A)$ in ${\bf {\bar 2}}$
representation have charges $(-,+)$. Since all matter fields are
represented by complex numbers(not matrices) in this case, the
fermionic and bosonic potentials in ABJM theory vanish (See eqs.
(\ref{ferV}) and (\ref{bosV})). Then the ABJM action with
U(1)$\times$U(1) gauge group is reduced to
\begin{align}\label{U1act}
S &=\int d^3x\,\left(
-D_\mu Z_A^\dagger D^{\mu} Z^A +
i\psi^{\dagger A} \gamma^\mu D_\mu \psi_A+
\frac{k}{4\pi}\,\epsilon^{\mu\nu\rho}A^-_\mu F^+_{\nu\rho}\right),\qquad
{\rm with}\,\,A=1,2,
\end{align}
where $D_\mu Z^A=\partial_\mu Z^A + iA^-_\mu Z^A$ and
$F^+_{\mu\nu}= \partial_\mu A^+_\nu - \partial_\nu A^+_\mu$.

Since the covariant derivative in (\ref{U1act}) depends only on
$A^-_\mu$, we can treat the field strength $F^+_{\mu\nu}$ as a
fundamental field and introduce a dual scalar field $\tau(x)$. In
order to do so, we have to add
\begin{align}\label{Lmult}
S=\frac{1}{4\pi}\int d^3 x\, \tau (x) \epsilon^{\mu\nu\rho}
\partial_\mu F^+_{\nu\rho}
\end{align}
to the action
(\ref{U1act})~\cite{Aharony:2008ug,Lambert:2008et,Distler:2008mk},
which represent the Bianchi identity for the gauge field
strength $F^+_{\mu\nu}$, after we integrate out the auxiliary scalar field
$\tau(x)$. Here $\tau (x)$ is $2\pi$-periodic due to the flux
quantization $\int d^3 x \epsilon^{\mu\nu\rho}\,\partial_\mu
F^+_{\nu\rho} = 4\pi n$ with integer $n$. Then the equation of motion of
$F^+_{\mu\nu}$ is given by
\begin{align}\label{eomF}
A^-_\mu = \frac{1}{k}\,\partial_\mu\tau.
\end{align}
The gauge transformation, $A^-_\mu \to A^-_\mu + \partial_\mu\Lambda$,
implies that $\tau(x)$ is transformed as $\tau \to \tau + k\Lambda$.
Though we fixed the gauge by taking $\tau=0$, we can still perform gauge
transformation with $\Lambda=\frac{2\pi}{k}$ from the periodic property
of $\tau$. This remaining  symmetry implies
\begin{align}\label{ofZ}
Z^A \sim e^{2\pi i/k}\, Z^A.
\end{align}
This means that the ${\cal N}=2$ minimal model (\ref{U1act}) with
U(1)$\times$U(1) gauge symmetry is reduced to the sigma model on
${\bf C}^2/{\bf Z}_k$ orbifold, while the original ABJM theory with
U(1)$\times$U(1) gauge group is reduced to the sigma model on ${\bf
C}^4/{\bf Z}_k$ orbifold. The form of action (\ref{U1act}) is
equivalent to that of ABJM action with U(1)$\times$U(1) gauge group,
though the number of fields in (\ref{U1act}) is half of that in ABJM
theory. So many of the physical properties of the minimal model with
U(1)$\times$U(1) gauge group are similarly with those of ABJM theory
with the same gauge group. Especially these two theories have the
same supersymmetry enhancement properties.

Due to the abelian properties of matter fields in the minimal model
with U(1)$\times$U(1) gauge group, the supersymmetry transformation
rules in (\ref{dN2susy}) are reduced to
\begin{align}\label{dN2U1}
\delta Z^A&= i\varepsilon^\dagger\epsilon^{AB}\psi_B,
\nn \\
\delta Z^\dagger_A&= i\epsilon_{AB}\psi^{\dagger B}\varepsilon,
\nn \\
\delta\psi_A&=\epsilon_{AB}D_\mu Z^B\gamma^\mu\varepsilon,
\nn \\
\delta\psi^{\dagger A}&=
-\varepsilon^\dagger\epsilon^{AB}\gamma^\mu D_\mu Z^\dagger_B,
\nn \\
\delta A_\mu&=\delta \hat A_\mu=
-\frac{2\pi}{k}\left(\varepsilon^\dagger\epsilon^{AB}\gamma_\mu
\psi_BZ^\dagger_A + \epsilon_{AB}Z^A\psi^{\dagger B}\gamma_\mu\varepsilon\right).
\end{align}
As we see in the supersymmetry transformation rules of
(\ref{dN2U1}), the variations of bosonic field $Z^A$ in ${\bf 2}$
representation of the global SU(2) is proportional to the fermionic
field $\psi_A$ in ${\bf {\bar 2}}$ representation. That is, the
${\cal N}=2$ supersymmetry (\ref{dN2U1}) relates the bosonic and
fermionic fields with different global SU(2) representations.
However, $Z^A$ and $\psi_A$ have the same ${\rm U}(1)_{A^-}$
charges.

Now we try to find the additional ${\cal N}=2$ supersymmetry in the
action (\ref{U1act}).  Differently from the ${\cal N}=2$ supersymmetry given
in (\ref{dN2U1}), the additional supersymmetry relates the bosonic
and fermionic fields with opposite ${\rm U}(1)_{A^-}$ charges
but with the same global SU(2) representations.
To compensate the ${\rm U}(1)_{A^-}$ charge differences,
we need to include some abelian operator with two units ${\rm U}(1)_{A^-}$
charge in the supersymmetry transformation rules of the additional
${\cal N}=2$ supersymmetry, as suggested in Ref.~\cite{Klebanov:2008vq}.

In order to find this kind of additional supersymmetry,
we insert a local operator in the expected
transformation rules, and find a condition for the operator to give
supersymmetric invariance of the action (\ref{U1act}).
At first, we consider the following ansatz for the supersymmetry
transformation rules,
\begin{align}\label{N2susy}
\delta_1 Z^\dagger_A&= i \tilde\varepsilon^\dagger \check T^* \psi_A,
\nn \\
\delta_1\psi^{\dagger A} &= \tilde\varepsilon^\dagger
\check T^*\gamma^\mu D^+_\mu Z^A,
\nn \\
\delta_1 A^-_\mu &=0,
\end{align}
where $\tilde\varepsilon^\dagger$ is the complex spinor
parameter\footnote{Throughout this paper, the complex spinor
parameter $\tilde\varepsilon$ and its complex conjugate
$\tilde\varepsilon^\dagger$ will be used to denote the additional
${\cal N}=2$ supersymmetry transformation rules.}, $D^{+}_\mu Z^A =
\partial_\mu Z^A + i A^-_\mu Z^A$, and we introduce a worldvolume
dependent operator $\check T^*=\check T^*(x)$ (complex conjugate of
$\check T$) to compensate the ${\rm U}(1)_{A^-}$ charge differences
in the relations (\ref{N2susy}). As will be explained later, $\check
T$ becomes the monopole operator. Here, we denoted the monopole
operator as $\check T$ in order to distinguish from the monopole
operator $T$ for the general U(N)$\times$ U(N) case. $\check T$ will
also appear in U(2)$\times$ U(2) case.
 We only analyzed half of the terms in obtaining the
supersymmetry transformation rules, since the remaining
transformation rules are easily obtained by taking complex conjugate
for (\ref{N2susy}). Applying the ansatz (\ref{N2susy}), we obtain
\begin{align}\label{susyvar1}
\delta_1{\cal L}=&
-i\tilde\varepsilon^\dagger\gamma^\mu D^+_\mu Z^A\gamma^\nu
\left(\partial_\nu \check T^*-2iA^-_\nu \check T^*\right)\psi_A
\nn \\
&-\frac{k}{4\pi} \epsilon^{\mu\nu\rho}A^-_\mu\partial_\nu
\left(\frac{4\pi}{k}\,\tilde\varepsilon^\dagger \check T^*
\gamma_\rho\psi_A Z^A -\delta_1 A^+_\rho\right),
\end{align}
up to   total derivative terms. From the relation (\ref{susyvar1}),
we see that the ${\cal N}=2$ minimal model (\ref{U1act}) is
invariant under the additional ${\cal N}=2$ supersymmetry
transformation (including the complex conjugate of
(\ref{N2susy})),
\begin{align} \label{susyvar2}
\delta Z^A&= i \check T\psi^{\dagger A}\tilde\varepsilon,
\nn \\
\delta Z^\dagger_A&= i \tilde\varepsilon^\dagger \check T^* \psi_A,
\nn \\
\delta\psi_A&= -\check T D^-_\mu Z^\dagger_A\gamma^\mu\tilde\varepsilon,
\nn \\
\delta\psi^{\dagger A} &= \tilde\varepsilon^\dagger
\check T^*\gamma^\mu D^+_\mu Z^A,
\nn \\
\delta A_\mu&=\delta \hat A_\mu =\frac{2\pi}{k}\left(\check T
\psi^{\dagger A}\gamma_\mu Z^\dagger_A\tilde\varepsilon +
\tilde\varepsilon^\dagger \check T^*\gamma_\mu \psi_A Z^A\right),
\end{align}
where $D^{-}_\mu Z_A^\dagger = \partial_\mu Z_A^\dagger
- i A^-_\mu Z_A^\dagger$ and $\check T$ satisfies the following differential
equation
\begin{align}\label{Teq}
\partial_\mu \check T + 2iA^-_\mu \check T = 0.
\end{align}
Here we should notice, however, that
the supersymmetry transformation rules (\ref{susyvar2}) are not satisfied
for all integer values of $k$, due to the orbifolding of matter fields
given in (\ref{ofZ}).

From now on, we solve the equation (\ref{Teq}) and try to
figure out the properties of the local operator $\check T(x)$.
Under the gauge transformation, $A_\mu^- \to A_\mu^- + \partial_\mu\Lambda$,
the matter field $(Z^A,\,\psi_A)$ with $+1$ ${\rm U}(1)_{A^-}$
charges, transform as $(Z^A,\,\psi_A)\to e^{-i\Lambda}(Z^A,\,\psi_A)$,
while $(Z^\dagger_A,\,\psi^{\dagger A})$ with $-1$ ${\rm U}(1)_{A^-}$
charges transform as $(Z^\dagger_A,\,\psi^{\dagger A}) \to e^{+i\Lambda}
(Z^\dagger_A,\,\psi^{\dagger A})$. Similarly from (\ref{Teq}), we can
easily see that $\check T$ transforms as
$\check T\to e^{-2i\Lambda} \check T$ under the same gauge transformation,
and so $\check T$ has $+2$ ${\rm U}(1)_{A^-}$ charge.
We can also check of the ${\rm U}(1)_{A^-}$ charge of $\check T$ by solving
the differential equation (\ref{Teq}) directly.
To guarantee the supersymmetry (\ref{dN2U1}) over the whole
worldvolume region, we consider the regular $\check T(x)$ only.
Then the equation (\ref{Teq}) implies that the ${\rm U}(1)_{A^-}$
gauge field $A_\mu^-$ is in pure gauge,
\begin{align}\label{AT}
A_\mu^- = \frac{i}{2} \partial_\mu \ln \check T,
\end{align}
which corresponds to the result (\ref{eomF}) by identifying
\begin{align}\label{tHop}
\check T(x)=\check T_0 e^{-2i \tau(x)/k}
\end{align}
with a constant $\check T_0$. Equivalently $\check T(x)$ can be expressed
by the path independent local Wilson line~\cite{Aharony:2008ug,Park:2008bk}
\begin{align}\label{wilT}
\check T(x) = \check T_0 e^{2i\int_x^\infty A_\mu^- dx^\mu},
\end{align}
which transforms as $\check T\to e^{-2i\Lambda} \check T$
under the gauge transformation
$A_\mu^-\to A_\mu^- + \partial_\mu\Lambda$. So we can also check that the
operator $\check T$ has $+2$ ${\rm U}(1)_{A^-}$ charge. It was argued that
attaching the local Wilson line to matter fields, we can
change the ${\rm U}(1)_{A^-}$ charges of the fields and
they are insensitive in the presence of the Wilson
line~\cite{Aharony:2008ug,Park:2008bk}.

Due to the Chern-Simons term in the action (\ref{U1act}), $\check T(x)$
becomes also the monopole operator (often called as 't Hooft
operator~\cite{'tHooft:1977hy}), which induces the magnetic flux from the
position $x$. In three dimensional Chern-Simons matter theory,
the Wilson line and 't Hooft operator are equivalent~\cite{Borokhov:2002ib},
as we have seen the equivalence between (\ref{wilT}) and (\ref{tHop})
which is the monopole operator.

Under a ${\bf Z}_k$ transformation for $\tau(x)$,
\begin{align}\label{ptau}
\tau(x) \to \tau(x) + 2\pi,
\end{align}
$\check T(x)$ given in (\ref{tHop}) transforms as
\begin{align}\label{Ttra}
\check T(x) \to e^{-4\pi i/k} \check T(x).
\end{align}
Therefore, the supersymmetry transformation rule is invariant under
${\bf Z}_k$ transformation only when $k=1,2$. It means
that the orbifold ($k\ge 3$) structure of the transverse space
breaks the supersymmetry invariance for the additional
supersymmetry (\ref{susyvar2}) of the action (\ref{U1act}).

\subsection{SU(2)$\times$SU(2) case}

The ${\cal N}=2$ minimal model (\ref{dact}) with SU(2)$\times$SU(2)
gauge group has additional ${\cal N}=2$ supersymmetry, in addition
to the ${\cal N}=2$ supersymmetry given in (\ref{dN2susy}).
Unlike the U(1)$\times$U(1) case which allows the additional ${\cal N}=2$
supersymmetry for $k=1,2$ cases only, for SU(2)$\times$SU(2) case
the additional ${\cal N}=2$ supersymmetry we will consider in this subsection
does not depend on the value of the Chern-Simons level $k$.
However, similarly to the case of U(1)$\times$U(1), the supersymmetric
variation of bosonic fields are proportional to fermionic fields
with same representations of the global SU(2) but different
gauge indices.

It turns out that the additional ${\cal N}=2$ supersymmetry
transformation rules of the matter fields in matrix notation are
given by
\begin{align}\label{SU2-1}
\delta Z^A &= i \tilde\psi^{\dagger A}\tilde\varepsilon,
\nn \\
\delta Z^\dagger_A &= i \tilde\varepsilon^\dagger\tilde\psi_A,
\nn \\
\delta\psi_A &= -D_\mu\tilde Z^\dagger_A\gamma^\mu\tilde\varepsilon -
\tilde N^\dagger_A\tilde\varepsilon,
\nn \\
\delta\psi^{\dagger A} &= \tilde\varepsilon^\dagger\gamma^\mu D_\mu\tilde Z^A
-\tilde\varepsilon^\dagger\tilde N^A,
\nn \\
\delta A_\mu&= \frac{2\pi}{k}\left(
\tilde\varepsilon^\dagger Z^A\gamma_\mu\tilde\psi_A + \tilde\psi^{\dagger A}
\gamma_\mu Z^\dagger_A\tilde\varepsilon\right),
\nn \\
\delta \hat A_\mu &= \frac{2\pi}{k}\left(
\tilde\varepsilon^\dagger \gamma_\mu\tilde\psi_A Z^A +
Z_A^\dagger\tilde\psi^{\dagger A}\gamma_\mu\tilde\varepsilon\right),
\end{align}
where $\tilde\varepsilon$ is the complex spinor parameter and
we define the fields with {\it tilde} notation as follows,
\begin{align}\label{dfntil}
\tilde Z^{A\hat a}_{~~~a}&\equiv \epsilon^{\hat a\hat b}\epsilon_{ab}
Z^{Ab}_{~~~\hat b},
\qquad
\tilde Z^{\dagger a}_{A~\hat a}\equiv \epsilon^{ab}\epsilon_{\hat a\hat b}
Z^{\dagger \hat b}_{A~b},
\nn \\
\tilde\psi^{\hat a}_{A a}&\equiv \epsilon^{\hat a\hat b}\epsilon_{ab}
\psi^b_{A\hat b}, \qquad
\tilde\psi^{\dagger A a}_{~~~\hat a}\equiv \epsilon^{ab}
\epsilon_{\hat a\hat b}
\psi^{\dagger A\hat b}_{~~~b},
\nn \\
\tilde N^{A\hat a}_{~~a}&\equiv \epsilon^{\hat a\hat b}\epsilon_{ab}
N^{Ab}_{~~~\hat b}, \qquad
\tilde N^{\dagger a}_{A~\hat a}\equiv
\epsilon^{ab}\epsilon_{\hat a\hat b} N^{\dagger\hat b}_{A~b}.
\end{align}
Here $\epsilon_{ab}$($\epsilon^{\hat a\hat b}$)
is the invariant tensor of the gauge group ${\rm SU}(2)_L$(${\rm SU}(2)_R$)
and we explicitly denote the gauge indices for definiteness.
In Appendix \ref{N2loc}, we verify the invariance
of the action (\ref{dact}) under the supersymmetry transformation
(\ref{SU2-1}).

Since there is no ${\rm U}(1)_{A^-}$ gauge symmetry in
SU(2)$\times$SU(2) case, the sypersymmetry transformation rules
(\ref{SU2-1}) do not include the monopole operators which do not
allow the additional ${\cal N}=2$ supersymmetry for $k>2$ cases. So
the minimal model  with SU(2)$\times$SU(2) gauge group has ${\cal
N}=4$ supersymmetry for arbitrary value of $k$.

\subsection{U(2)$\times$U(2) with both cases united}\label{U2U2section}

We investigated the supersymmetry enhancement of the ${\cal N}=2$ minimal
model (\ref{dact}) in the previous subsections. In U(1)$\times$U(1)
case, the ${\cal N}=2$ supersymmetry (\ref{N2susy}) is enhanced to
${\cal N}=4$ supersymmetry at $k=1,2$ only. In
SU(2)$\times$SU(2) case, however, the ${\cal N}=2$ supersymmetry
(\ref{N2susy}) is enhanced to ${\cal N}=4$ supersymmetry regardless
the value of $k$.

Similarly to the cases of U(1)$\times$U(1) and SU(2)$\times$SU(2),
we find the additional ${\cal N}=2$ supersymmetry transformation
rules for U(2)$\times$U(2) gauge group as follows
\begin{align}\label{U2-1}
\delta Z^A &= i \tilde\psi^{\dagger A}\check T\tilde\varepsilon,
\nn \\
\delta Z^\dagger_A &= i \check T^*\tilde\varepsilon^\dagger\tilde\psi_A,
\nn \\
\delta\psi_A &= -D_\mu\tilde Z^\dagger_A\gamma^\mu \check T
\tilde\varepsilon -\tilde N^\dagger_A \check T\tilde\varepsilon,
\nn \\
\delta\psi^{\dagger A} &= \check T^*\tilde\varepsilon^\dagger\gamma^\mu
D_\mu\tilde Z^A -\check T^*\tilde\varepsilon^\dagger\tilde N^A,
\nn \\
\delta A_\mu &= \frac{2\pi}{k}\left(
\check T^*\tilde\varepsilon^\dagger Z^A\gamma_\mu\tilde\psi_A
+ \tilde\psi^{\dagger A} \gamma_\mu Z^\dagger_A \check T
\tilde\varepsilon\right),
\nn \\
\delta \hat A_\mu &= \frac{2\pi}{k}\left(
\check T^*\tilde\varepsilon^\dagger \gamma_\mu\tilde\psi_A Z^A +
Z_A^\dagger\tilde\psi^{\dagger A}\gamma_\mu \check T
\tilde\varepsilon\right),
\end{align}
where $\tilde N^A$ was defined in (\ref{dfntil}) and
$\check T$ is the abelian monopole operator given in (\ref{tHop})
or (\ref{wilT}) with the ${\rm U}(1)_{A^-}$ gauge field,
$A^-_\mu = {\rm tr} A_\mu-{\rm tr} \hat A_\mu$.

From (\ref{U2-1}), we can obtain the supersymmetry transformation
(\ref{SU2-1}) of SU(2)$\times$SU(2) case by removing the monopole
operator $\check T$. And also, we can obtain the supersymmetry
transformation (\ref{susyvar2}) of U(1)$\times$U(1) case from
(\ref{U2-1}) by regarding all fields in (\ref{U2-1}) as complex numbers
without gauge indices. The reason is as
follows. Dividing the gauge fields, $A_\mu$ and $\hat A_\mu$,  into
the trace part and traceless part, we can also decompose the action
(\ref{dact}) with U(2)$\times$U(2) gauge group into two parts with
U(1)$\times$U(1) and SU(2)$\times$SU(2) gauge groups
respectively. The gauge fields can be rewritten as
\begin{align}\nn
A_\mu = B_\mu + C_\mu,\qquad \hat A_\mu = \hat B_\mu + \hat C_\mu,
\end{align}
where $B_\mu = {\rm tr} (A_\mu) \frac{{\bf 1}}{2}$ and
$\hat B_\mu = {\rm tr} (\hat A_\mu) \frac{{\bf 1}}{2}$
with 2$\times$2 identity matrix ${\bf 1}$.
Using the trace property, we can rewrite
the kinetic and Chern-Simons terms in (\ref{dact}) as
\begin{align}\label{U2dact}
{\cal L}_0&= {\rm tr}\left(- D_\mu Z_A^\dagger  D^\mu Z^A
+ i \psi^{\dagger A}\gamma^\mu  D_\mu\psi_A\right),
\nn \\
{\cal L}_{{\rm CS}} &= \frac{k}{4\pi}\left(
B_\mu\partial_\nu B_\rho - \hat B_\mu\partial_\nu \hat B_\rho\right)
\nn \\
&~
+\frac{k}{4\pi}\,\epsilon^{\mu\nu\rho}\,{\rm tr}
\left(C_\mu \partial_\nu C_\rho +\frac{2i}{3}C_\mu C_\nu C_\rho
- \hat{C}_\mu \partial_\nu \hat{C}_\rho
-\frac{2i}{3}\hat{C}_\mu \hat{C}_\nu \hat{C}_\rho \right),
\end{align}
where the covariant derivative is decomposed by
\begin{align}\label{covD2}
D_\mu Z^A = \partial Z^A + i (B_\mu - \hat B_\mu ) Z^A +
i C_\mu Z^A - i Z^A \hat C_\mu.
\end{align}
Since there is no potential for U(1)$\times$U(1) case, we can think
that the potentials in U(2)$\times$U(2) case are decomposed into
U(1)$\times$U(1) part and SU(2)$\times$SU(2) part already. From
these reasons, we can read the supersymmetry transformation rules
(\ref{susyvar2}) and (\ref{SU2-1}) from (\ref{U2-1}), and vice
versa.

Since the supersymmetry transformation rules (\ref{U2-1}) include
the monopole operator $\check T$, according to the discussion of subsection
\ref{U1U1section} the supersymmetry of the ${\cal N}=2$ minimal
model with U(2)$\times$U(2) gauge group is enhanced to ${\cal N}=4$
for $k=1,2$ cases only.

\section{Supersymmetry Enhancement of the ABJM Theory with U(2)$\times$U(2)
Gauge Group}

The ABJM action with U($N$)$\times$U($N$)
gauge group at Chern-Simons level ($k$,$-k$) in SU(4) invariant form
is given by
\begin{align}\label{ABJMact}
S =\int d^3x\,\left({\cal L}_0 + {\cal L}_{{\rm CS}} -V_{{\rm ferm}}
-V_{{\rm bos}} \right)
\end{align}
with
\begin{align}
{\cal L}_0 &= {\rm tr}\left(-D_\mu Y_A^\dagger D^\mu Y^A +
i\psi^{\dagger A} \gamma^\mu D_\mu \psi_A\right),
\label{ABJMkin} \\
{\cal L}_{{\rm CS}} &= \frac{k}{4\pi}\,\epsilon^{\mu\nu\rho}\,{\rm tr}
\left(A_\mu \partial_\nu A_\rho +\frac{2i}{3}A_\mu A_\nu A_\rho
- \hat{A}_\mu \partial_\nu \hat{A}_\rho
-\frac{2i}{3}\hat{A}_\mu \hat{A}_\nu \hat{A}_\rho \right),
\label{ABJMcs} \\
V_{{\rm ferm}} &= \frac{2\pi i}k{\rm tr}\Big( Y_A^\dagger Y^A\psi^{\dagger
B}\psi_B -Y^A Y_A^\dagger\psi_B \psi^{\dagger B}
+2Y^AY_B^\dagger\psi_A\psi^{\dagger B} -2Y_A^\dagger
Y^B\psi^{\dagger A}\psi_B
\label{ferV} \\
&\hskip 1.7cm  +\epsilon^{ABCD}Y^\dagger_A\psi_BY^\dagger_C\psi_D
-\epsilon_{ABCD}Y^A\psi^{\dagger B}Y^C\psi^{\dagger D} \Big),
\nn \\
V_{{\rm bos}} &=-\frac{4\pi^2}{3k^2}{\rm tr}\Big(
Y^\dagger_AY^AY^\dagger_BY^BY^\dagger_CY^C
+Y^AY^\dagger_AY^BY^\dagger_BY^CY^\dagger_C
+4Y^\dagger_AY^BY^\dagger_CY^AY^\dagger_BY^C
\label{bosV} \\
&\hskip 2cm -6Y^AY^\dagger_BY^BY^\dagger_AY^CY^\dagger_C \Big).\nn
\end{align}
Here the complex scalars $Y^A$, ($A=1,\cdots,4$), the fermions $\psi_A$
are in bifundamental representation, and the covariant derivatives are same
as (\ref{covD}).

The ABJM action (\ref{ABJMact}) is invariant under ${\cal N}=6$
supersymmetry transformation,
\begin{align}
&\delta Y^A = i \omega^{AB}\psi_B,
\nn \\
&\delta Y^{\dagger}_A = i \psi^{\dagger B}\omega_{AB},
\nn \\
&\delta\psi_A = \gamma^\mu\omega_{AB}D_\mu Y^B
+\frac{2\pi}k\omega_{AB}(Y^BY^\dagger_CY^C -Y^CY^\dagger_CY^B)
+\frac{4\pi}k\omega_{BC}Y^BY^\dagger_AY^C,
\nn \\
&\delta\psi^{\dagger\,A}= -D_\mu Y^\dagger_B \omega^{AB}\gamma^\mu
+\frac{2\pi}k\omega^{AB}(Y^\dagger_CY^CY^\dagger_B
-Y^\dagger_BY^CY^\dagger_C)
-\frac{4\pi}k\omega^{BC}Y^\dagger_BY^AY^\dagger_C,
\nn \\
&\delta A_\mu = -\frac{2\pi}{k}(\omega^{AB}Y^{\dagger}_A
\gamma_\mu\psi_B + Y^A \psi^{\dagger B}\gamma_\mu\omega_{AB}),
\nn \\
&\delta \hat{A}_\mu =
-\frac{2\pi}{k}(\omega^{AB}Y^\dagger_A\gamma_\mu\psi_B
+\psi^{\dagger B}\gamma_\mu Y^A\omega_{AB}),
\label{N6part}
\end{align}
where $\omega^{AB}=-\omega^{BA}=(\omega_{AB})^*=
\frac{1}{2}\,\epsilon^{ABCD}\omega_{CD}$.

\subsection{U(2)$\times$U(2) case}

It was conjectured that the ABJM theory with U($N$)$\times$U($N$)
gauge group has the additional ${\cal N}=2$ supersymmetry at
$k=1,2$, in addition to ${\cal N}=6$ supersymmetry written
in (\ref{N6part})~\cite{Aharony:2008ug}. Since the equivalence between
the SU(2)$\times$SU(2) ABJM theory and BLG theory which has ${\cal
N}=8$ supersymmetry was known in Ref.~\cite{Aharony:2008ug} already,
we know that the ABJM theory with SU(2)$\times$SU(2) gauge group has
${\cal N}=8$ supersymmetry without dependence of $k$, i.e. there is
${\cal N}=2$ supersymmetry enhancement in SU(2)$\times$SU(2) case though the
supersymmetry transformation rules for component fields were not
known up to now. In Appendix \ref{B1}, we verify the supersymmetry
invariance of the ABJM theory with SU(2)$\times$SU(2) gauge group
and find the corresponding additional ${\cal N}=2$ supersymmetry
transformation rules. On the other hand, for U(1)$\times$U(1) case,
the ${\cal N}=2$ minimal model and ABJM theory have the same kinetic
and Chern-Simons terms. So they have the same supersymmetric
behaviors for the additional ${\cal N}=2$ supersymmetry, though the
numbers of matter fields are different. Actually the additional
${\cal N}=2$ supersymmetry given in (\ref{susyvar2}) for
U(1)$\times$U(1) case does not dependent on the number of matter
fields.

As we discussed in the subsection \ref{U2U2section}, in order to
obtain the additional supersymmetry transformation rules of U(2)$\times$U(2)
we can combine the results of U(1)$\times$U(1) and SU(2)$\times$SU(2) cases.
The results are as follows:
\begin{align}\label{U2-2}
\delta Y^A &= i \check T\tilde\psi^{\dagger A}\tilde\varepsilon,
\nn \\
\delta Y^\dagger_A &= i \tilde\varepsilon^\dagger \check T^*\tilde\psi_A,
\nn \\
\delta\psi_A &= -\check T D_\mu\tilde
Y^\dagger_A\gamma^\mu\tilde\varepsilon - \check T\tilde
N^\dagger_A\tilde\varepsilon - \frac{4\pi}{3k}\,\check T^*
\tilde\varepsilon^\dagger \epsilon_{ABCD}Y^B\tilde Y^C Y^D,
\nn \\
\delta\psi^{\dagger A} &= \check T^*\tilde\varepsilon^\dagger \gamma^\mu
D_\mu\tilde Y^A -\check T^*\tilde\varepsilon^\dagger \tilde N^A +
\frac{4\pi}{3k}\, \check T\epsilon^{ABCD} Y^\dagger_B
\tilde Y^\dagger_CY^\dagger_D
\tilde\varepsilon,
\nn \\
\delta A_\mu &= \frac{2\pi}{k}\left(\check T^*
\tilde\varepsilon^\dagger Y^A\gamma_\mu\tilde\psi_A
+ \check T\tilde\psi^{\dagger A}
\gamma_\mu Y^\dagger_A\tilde\varepsilon \right),
\nn \\
\delta \hat A_\mu &= \frac{2\pi}{k}\left(\check T^*
\tilde\varepsilon^\dagger \gamma_\mu\tilde\psi_A Y^A +
\check T Y_A^\dagger\tilde\psi^{\dagger A}
\gamma_\mu\tilde\varepsilon \right),
\end{align}
where $\tilde Y^A$, $\tilde \psi_A$, and $\tilde N^A$ are defined
in (\ref{dfntil}) by replacing $Z$ with $Y$.

Similarly to the case of the ${\cal N}=2$ minimal model with
U(2)$\times$U(2) gauge group discussed in the subsection
\ref{U2U2section}, the supersymmetry of the ${\cal N}=6$ ABJM theory
with U(2)$\times$U(2) gauge group is enhanced to ${\cal N}=8$ for $k=1,2$
cases due to the presence of the abelian monopole operator $\check T$.

\subsection{Comments on U($N$)$\times$U($N$) case}

Now we try to extend our results to SU($N$)$\times$SU($N$)
case\footnote{Except for U(1)$\times$U(1) factor, there is no
difference between U($N$)$\times$U($N$) and SU($N$)$\times$SU($N$)
cases in obtaining
 supersymmetry transformation rules. As we have seen in the subsection
\ref{U2U2section}, we can consider the U(1)$\times$U(1) part
separately. So concentrating on SU($N$)$\times$SU($N$) case can
cover the U($N$)$\times$U($N$) case also.}. Here, we only give a
sketchy of the whole procedure whose explicit construction for the
supersymmetry transformation rules will  complete of the
supersymmetry enhancement in ABJM theory. The details will be
published elsewhere.

As we did in Appendix to prove the supersymmetry invariance of the
given Lagrangian in Chern-Simons matter theory, we start from the
following variations for the scalars and fermions:
\begin{align}\label{gauind}
\delta_1 Y^{\dagger\hat a}_{A~ a}&= i \tilde\varepsilon^\dagger
T^{\hat a\hat b}_{ab} \psi^{b}_{A\hat b},
\nn \\
\delta_1\psi^{\dagger A\hat a}_{~~~ a} &=
\tilde\varepsilon^\dagger T^{\hat a\hat b}_{ab}
\gamma^\mu D_\mu Y^{Ab}_{~~~\hat b},
\end{align}
where we denoted the gauge indices for concreteness.
Since all the gauge groups except for SU(2)$\times$SU(2) include the
U(1)$\times$U(1) part, the local operator $T$ transforms as
$T\to e^{-4\pi i/k} T$
under the orbifold transformation for the generic gauge group.
Therefore, the supersymmetry transformation (\ref{gauind})
is satisfied in the cases $k=1,2$ only.
Here $D_\mu Y^{Aa}_{~~~\hat a}$, $(a,b,... = 1,2,...N; \hat a,\hat b,...
= 1,2,... N)$, and its Hermitian conjugate are given by
\begin{align}
D_\mu Y^{Aa}_{~~~\hat a}&= \partial_\mu Y^{Aa}_{~~~\hat a} + i A^a_{\mu b}
Y^{Ab}_{~~~\hat a}- i Y^{Aa}_{~~~\hat b}\hat A^{\hat b}_{\mu\hat a},
\nn \\
D_\mu Y^{\dagger \hat a}_{A~ a}&=\partial_\mu Y^{\dagger \hat a}_{A~ a}
+ i \hat A^{\hat a}_{\mu \hat b} Y^{\dagger \hat b}_{A~ a}
- i Y^{\dagger \hat a}_{A~b} A^{b}_{\mu a}.
\end{align}
The variation of ${\cal L}_0$ in (\ref{ABJMkin}) in matrix notation
is given by
\begin{align}\label{d1L-3}
\delta_1 {\cal L}_0 &= {\rm tr}\Big(-D_\mu\delta_1 Y_A^\dagger D^\mu
Y^A + i \delta_1\psi^{\dagger A}\gamma^\mu D_\mu\psi_A\Big)
\nn \\
&={\rm tr}\Big(-i\tilde\varepsilon^\dagger D_\mu(T\psi_A) D^\mu Y^A
+i\tilde\varepsilon^\dagger TD_\mu Y^A \gamma^\mu\gamma^\nu
D_\nu\psi_A\Big)
\nn \\
&={\rm tr}\Big(-i\tilde\varepsilon^\dagger\gamma^\nu\gamma^\mu D_\mu
T\psi_AD_\nu Y^A
-\frac{1}{2}\tilde\varepsilon^\dagger\epsilon^{\mu\nu\rho}
TF_{\mu\nu}Y^A\gamma_\rho\psi_A +
\frac{1}{2}\tilde\varepsilon^\dagger\epsilon^{\mu\nu\rho} TY^A\hat
F_{\mu\nu}\gamma_\rho\psi_A\Big).
\end{align}
In the final step of (\ref{d1L-3}) we integrated by part, dropped the total
derivative term, and used the following relation,
\begin{align}\label{DTp-1}
D_\mu (T\psi_A) = (D_\mu T)\psi_A + TD_\mu \psi_A.
\end{align}
Denoting the gauge indices we can express the left hand side of (\ref{DTp-1})
as
\begin{align}\label{DTp-2}
D_\mu (T\psi_A)^{\hat a}_{~ a}
&=(\partial_\mu T^{\hat a\hat b}_{ab})\psi^b_{A\hat b} +
T^{\hat a\hat b}_{ab}(\partial_\mu \psi^b_{A\hat b})
+ i \hat A^{\hat a}_{\mu\hat b} T^{\hat b\hat c}_{ac}\psi^c_{A\hat c}
-i T^{\hat a\hat c}_{bc}\psi^c_{A\hat c} A^b_{\mu a}
\end{align}
and the right hand side of (\ref{DTp-1}) as
\begin{align}\label{DTp-3}
\left((D_\mu T)\psi_A\right)^{\hat a}_{~a}
+ \left(TD_\mu \psi_A\right)^{\hat a}_{~a}
&= (D_\mu T)^{\hat a\hat b}_{ab} \psi^b_{A\hat b} + T^{\hat a\hat
b}_{ab}\left( \partial_\mu\psi^b_{A\hat b} +i A^b_{\mu
c}\psi^c_{A\hat b} -i \psi^b_{A\hat c}\hat A^{\hat c}_{\mu\hat
b}\right).
\end{align}
Combining (\ref{DTp-1}), (\ref{DTp-2}), and (\ref{DTp-3}), we obtain
\begin{align}\label{DTp-4}
(D_\mu T)^{\hat a\hat b}_{ab} \psi^b_{A\hat b}
&=\left[\partial_\mu T^{\hat a\hat b}_{ab} +i\hat A^{\hat
a}_{\mu\hat c} T^{\hat c\hat b}_{ab} + i \hat A^{\hat b}_{\mu\hat
c}T^{\hat a\hat c}_{ab} -i T^{\hat a\hat b}_{cb}A^c_{\mu a} -i
T^{\hat a\hat b}_{ac}A^c_{\mu b} \right]\psi^b_{A\hat b}.
\end{align}
On the other hand, the variation of ${\cal L}_{{\rm CS}}$ in (\ref{ABJMcs})
is given by
\begin{align}\label{CSterm}
\delta_A{\cal L}_{{\rm CS}} &= \frac{k}{4\pi}
\epsilon^{\mu\nu\rho}{\rm tr} \left(F_{\mu\nu}\delta A_\rho - \delta\hat A_\rho
\hat F_{\mu\nu}\right).
\end{align}

Adding (\ref{d1L-3}) and (\ref{CSterm}) we obtain
\begin{align}\label{d1A}
\delta_1 {\cal L}_0 + \delta_A{\cal L}_{{\rm CS}}&=
-i\tilde\varepsilon^\dagger\gamma^\nu\gamma^\mu D_\mu T^{\hat a\hat
b}_{ab} \psi^b_{A\hat b}D_\nu Y^{Aa}_{~~~\hat a}
\nn \\
&~~-\frac{1}{2}\tilde\varepsilon^\dagger\epsilon^{\mu\nu\rho}
T^{\hat a\hat b}_{ab} F^b_{\mu\nu c}Y^{Ac}_{~~~\hat b}\gamma_\rho
\psi^a_{A\hat a} +
\frac{1}{2}\tilde\varepsilon^\dagger\epsilon^{\mu\nu\rho}
T^{\hat a\hat b}_{ab}Y^{Ab}_{~~~\hat c}\hat F^{\hat c}_{\mu\nu\hat b}
\gamma_\rho\psi^a_{A\hat a}
\nn \\
&~~+ \frac{k}{4\pi}\epsilon^{\mu\nu\rho}F^b_{\mu\nu c}\delta A^c_{\rho b}
-\frac{k}{4\pi}\epsilon^{\mu\nu\rho}\delta \hat A^{\hat b}_{\rho\hat c}
\hat F^{\hat c}_{\mu\nu\hat b}.
\end{align}
As we did in the relations (\ref{dL1}) and (\ref{dL1-2}) which
appeared in proving the supersymmetry invariance of the Chern-Simons
matter theories, we first impose the vanishing of the right hand
side of (\ref{d1A}). Then we obtain the variations for the gauge
field and a condition for $T^{\hat a\hat b}_{ab}$ as follows:
\begin{align}
&\delta A^a_{\mu b}= \frac{2\pi}{k}\tilde\varepsilon^\dagger
Y^{A a}_{~~~\hat a}\gamma_\mu T^{\hat a\hat b}_{bc}\psi^c_{A\hat b},
\nn\\
&\delta\hat A^{\hat a}_{\mu\hat b}=
\frac{2\pi}{k}\tilde\varepsilon^\dagger
\gamma_\mu T^{\hat a\hat c}_{ab}\psi^b_{A\hat c}Y^{A a}_{~~~\hat b},
\label{dhAmu}\\
&(D_\mu T)^{\hat a\hat b}_{ab} =
\partial_\mu T^{\hat a\hat b}_{ab} +i\hat A^{\hat
a}_{\mu\hat c} T^{\hat c\hat b}_{ab} + i \hat A^{\hat b}_{\mu\hat
c}T^{\hat a\hat c}_{ab} -i T^{\hat a\hat b}_{cb}A^c_{\mu a} -i
T^{\hat a\hat b}_{ac}A^c_{\mu b} =0.\label{DTeq0}
\end{align}

The supersymmetry transformation rules for SU($N$)$\times$SU($N$)
case are read from (\ref{gauind}) and (\ref{dhAmu}). Though
$Y^3$-terms in the variation of fermion fields, which are originated
from the variation of the potentials are not available, the rules
(\ref{gauind}) and (\ref{dhAmu}) will be very useful in finding the
complete supersymmetry transformation rules in
SU($N$)$\times$SU($N$) case.

\section{Conclusion}

We investigated the supersymmetry enhancement behaviors of the ABJM
theory. We found the additional ${\cal N}=2$ supersymmetry
explicitly for U(1)$\times$U(1) and  U(2)$\times$U(2) cases at
$k=1,2$, by introducing the local operator $\check T$ which is known
as monopole operator. In obtaining the additional supersymmetry
transformation rules, we started from the verification of the
supersymmetric invariance for the minimal model which has the same
supersymmetry enhancement properties as those of ABJM theory. The
minimal model is a ${\cal N}=2$ supersymmetric Chern-Simons matter
theory with U(1)$_R$ symmetry and has the same forms of the kinetic
terms for scalars and fermions, the Chern-Simons terms. The matter
field part is composed of two complex scalars and fermions.  We
found the explicit supersymmetry transformation rules for ${\cal
N}=2$ supersymmetry coming from the global symmetry and the
additional ${\cal N}=2$ supersymmetry originated from the gauge part
for U(1)$\times$U(1) and U(2)$\times$U(2) at $k=1,2$. That is, the
minimal model has ${\cal N}=4$ supersymmetry at $k=1,2$.

 The procedure of the minimal
model can be repeated to ABJM theory, and we proved the conjecture
for the supersymmetry enhancement  for U(1)$\times$U(1) and
U(2)$\times$U(2) cases.  We  explicitly obtained the additional
${\cal N}=2$ supersymmetry transformation rules by using the monopole
operator and showed ${\cal N}=8$ supersymmetry of ABJM theory.
We also studied the additional ${\cal N}=2$ supersymmetry for the
generic gauge group U($N$)$\times$U($N$) case without the contribution
from the potentials of ABJM theory, and derived a
condition for the monopole operator $T$, which satisfies first order
coupled differential equation. Since $T$ has four indices, solving
the coupled differential equations is nontrivial and we were not
able to complete  the computation of $Y^3$-terms in the variation of
the fermion fields, which seems to require a more lengthy
calculation. But we believe that the variations for the potential
part give some additional condition for $T$ and we can reduce the
degrees of freedom for $T$ considerably.

As the extensions of the works related to the ${\cal N}=6$ supersymmetry
of ABJM theory, there are several directions we can consider
by using the additional ${\cal N}=2$ supersymmetry (\ref{U2-2}),
for instance, supersymmetry preserving mass
deformation~\cite{Gomis:2008cv, Hosomichi:2008qk, Hosomichi:2008jb,
Gomis:2008vc}, various soliton solutions~\cite{Hosomichi:2008qk,
Krishnan:2008zm, Terashima:2008sy, Jeon:2008zj, Kim:2008cp, Arai:2008kv,
Kim:2009ny}.

We conclude with a final remark.
Recently the non-relativistic limit
of ABJM theory was obtained in Refs.~\cite{Nakayama:2009cz,Lee:2009mm}. In
the non-relativistic theories the ${\cal N}=6$ part of supersymmetry
of ABJM theory was reduced to the kinematical, dynamical, and
conformal charges. It is also interesting to consider the
non-relativistic reduction for the additional ${\cal N}=2$ part of
supersymmetry of ABJM theory~\cite{KOSS}.

\section*{Acknowledgements}
We are grateful to Hiroaki Nakajima for his early stage of collaboration
and also for many helpful discussions.
We also thank Yoonbai Kim,  Cheonsoo Park, and Sang-Heon Yi for helpful discussions and comments.
This work  is supported by the Science Research Center Program
of the Korea Science and Engineering Foundation through the Center
for Quantum Spacetime(CQUeST) of Sogang University with grant No.
R11-2005-021 (P.O.) and the Korea Research Foundation(KRF)
grant funded by the Korea government(MEST) (No. 2009-0073775) (O.K.).

\appendix

\section{Supersymmetry of the ${\cal N}=2$ Minimal Model}
\subsection{Verification of (\ref{dN2susy})}\label{N2global}

Let us check the ${\cal N}=2$ supersymmetry (\ref{dN2susy}) of the action
(\ref{dact}). As we did in the U(1)$\times$U(1) case in the subsection
\ref{U1U1section}, we only check half of terms. The remaining half are
complex conjugate of them. From now on, we drop total derivatives
in all calculational procedures.

A supersymmetric variation of the kinetic term ${\cal L}_0$ (\ref{dakin})
is given by
\begin{align}\label{d1L}
\delta_1 {\cal L}_0 &= {\rm tr}\left( - D_\mu Z_A^\dagger D^\mu \delta_1
Z^A + i \delta_1 \psi^{\dagger B} \gamma^\nu D_\nu\psi_B\right)
\nn \\
&= {\rm tr}\left[ \frac{1}{2} \epsilon^{\mu\nu\rho} F_{\mu\nu}
\left(\varepsilon^\dagger \epsilon^{AB}\gamma_\rho \psi_B Z_A^\dagger\right)
-\frac{1}{2}\epsilon^{\mu\nu\rho}\left(\varepsilon^\dagger
\epsilon^{AB} Z_A^\dagger \gamma_\rho\psi_B\right)\hat F_{\mu\nu}\right],
\end{align}
where $F_{\mu\nu}$ and $\hat F_{\mu\nu}$ are gauge field strengths of
$A_\mu$ and $\hat A_{\mu}$ respectively, and
\begin{align}\label{dzpsi}
\delta_1 Z^A &= i\varepsilon^\dagger\epsilon^{AB}\psi_B,
\nn \\
\delta_1\psi^{\dagger B} &= \varepsilon^\dagger\epsilon^{AB}D_\mu Z_A^\dagger
\gamma^\mu.
\end{align}
In the last step of (\ref{d1L}), we used integration by part and
\begin{align}\nn
[D_\mu,\, D_\nu] Z_A^\dagger = i \hat F_{\mu\nu} Z_A^\dagger - i Z_A^\dagger
F_{\mu\nu}.
\end{align}
We add $\delta_1 {\cal L}_0$ in (\ref{d1L}) to supersymmetric variations of
gauge fields in the Chern-Simons terms,
\begin{align}\nn
\delta_A {\cal L}_{{\rm CS}} = {\rm tr} \left(
\frac{k}{4\pi} \epsilon^{\mu\nu\rho} F_{\mu\nu} \delta A_\rho
- \frac{k}{4\pi} \epsilon^{\mu\nu\rho} \delta\hat A_\rho \hat F_{\mu\nu}
\right).
\end{align}
If we set the supersymmetric variations of gauge fields as follows,
\begin{align}\label{dAm}
\delta A_\mu &= -\frac{2\pi}{k} \varepsilon^\dagger\epsilon^{AB} \gamma_\mu
\psi_B Z_A^\dagger,
\nn \\
\delta \hat A_\mu &= -\frac{2\pi}{k}\varepsilon^\dagger \epsilon^{AB}
Z_A^\dagger \gamma_\mu\psi_B,
\end{align}
we obtain
\begin{align}\label{dL1}
\delta_1 {\cal L}_0 + \delta_A {\cal L}_{{\rm CS}}=0.
\end{align}

Now we consider the variation of the kinetic part ${\cal L}_0$ originated
form the variations of gauge fields,
\begin{align}\label{dAL}
\delta_A {\cal L}_0 &= {\rm tr} \left[ i Z_C^\dagger \delta A^\mu
D_\mu Z^C - i D_\mu Z_C^\dagger \delta A^\mu Z^C - \psi^{\dagger C}
\gamma^\mu \delta A_\mu \psi_C \right.
\nn \\
&\left.\hskip 1cm - i \delta\hat A^\mu Z_C^\dagger D_\mu Z^C
+ i D_\mu Z_C^\dagger Z^C
\delta\hat A^\mu + \psi^{\dagger C}\gamma^\mu\psi_C\delta\hat A_\mu\right].
\end{align}
Plugging (\ref{dAm}) into (\ref{dAL}), we obtain, after some algebra,
the following relation,
\begin{align}\label{dL2}
\delta_A {\cal L}_0 &= - {\rm tr}\left( i\delta_2\psi^{\dagger B}
\gamma^\mu D_\mu\psi_B\right) + \delta_1 V_{{\rm ferm}}
\nn \\
&= - \delta_2 {\cal L}_0 + \delta_1 V_{{\rm ferm}},
\end{align}
where
\begin{align}
\delta_2 \psi^{\dagger B} &= -\frac{2\pi}{k} \varepsilon^\dagger\epsilon^{AB}
\left( Z_A^\dagger Z^C Z_C^\dagger - Z_C^\dagger Z^C Z_A^\dagger\right),
\label{d2psi}\\
V_{{\rm ferm}} &= \frac{2\pi i}{k} {\rm tr}
\left(Z_A^\dagger Z^A \psi^{\dagger B}\psi_B
- Z^A Z_A^\dagger\psi_B\psi^{\dagger B}
+ 2Z^A Z_B^\dagger\psi_A\psi^{\dagger B} -2 Z_A^\dagger Z^B\psi^{\dagger A}
\psi_B\right).\label{Vferm}
\end{align}

As a next step, we consider the variation of $V_{{\rm ferm}}$ originated
from $\delta_2\psi^{\dagger B}$ given in (\ref{d2psi}),
\begin{align}\label{dL3}
\delta_2 V_{{\rm ferm}} &= \frac{2\pi i}{k} {\rm tr} \left(
\delta_2 \psi^{\dagger B}\psi_B Z_D^\dagger Z^D - \delta_2\psi^{\dagger B}
Z^D Z_D^\dagger\psi_B +2\delta_2\psi^{\dagger B} Z^D Z_B^\dagger\psi_D
-2\delta_2\psi^{\dagger B}\psi_D Z_B^\dagger Z^D\right)
\nn \\
&= - \delta_1 V_{{\rm bos}},
\end{align}
where
\begin{align}\label{Vbos}
V_{{\rm bos}} = \frac{4\pi}{k^2} {\rm tr}\left(
Z_A^\dagger Z^A Z_B^\dagger Z^B Z_C^\dagger Z^C
+Z^AZ_A^\dagger Z^B Z_B^\dagger Z^CZ_C^\dagger - 2Z^AZ_B^\dagger Z^BZ_A^\dagger
Z^CZ_C^\dagger\right).
\end{align}
Since there is no fermion field in the expression of $V_{{\rm bos}}$,
we see that $\delta_1 V_{{\rm bos}}=(\delta_1 + \delta_2)V_{{\rm bos}}
= \delta V_{{\rm bos}}$.

From the relations, (\ref{dzpsi}), (\ref{dAm}), and (\ref{d2psi}),
we obtain the total supersymmetric variations for component fields,
\begin{align}
\delta Z^A &= \delta_1 Z^A,
\nn \\
\delta \psi^{\dagger A} &= \delta_1\psi^{\dagger A}
+ \delta_2\psi^{\dagger A}.\nn
\end{align}
Then the total supersymmetric variation for the Lagrangian in (\ref{dact})
can be obtained from the equations, (\ref{dL1}), (\ref{dL2}),
and (\ref{dL3}), as follows,
\begin{align}\nn
\delta ({\cal L}_0 + {\cal L}_{{\rm CS}} - V_{{\rm ferm}}
- V_{{\rm bos}}) = \delta {\cal L} = 0.
\end{align}
Adding the complex conjugate parts of the supersymmetry transformation
rules, we prove that the action (\ref{dact}) is invariant under the
${\cal N}=2$ supersymmetry given in (\ref{dN2susy}).

\subsection{Verification of (\ref{SU2-1})}\label{N2loc}

As we did in the previous subsection, we start form a variation of the
kinetic part ${\cal L}_0$ given in (\ref{dakin}),
\begin{align}\label{d1L-2}
\delta_1 {\cal L}_0 &= {\rm tr}\left( - D_\mu \delta_1 Z_A^\dagger
D^\mu Z^A + i\delta_1\psi^{\dagger A} \gamma^\nu D_\nu\psi_A\right)
\nn \\
&= {\rm tr}\left(-\frac{1}{2} \epsilon^{\mu\nu\rho} F_{\mu\nu}
(\tilde\varepsilon^\dagger Y^A\gamma_\rho\tilde\psi_A)
+\frac{1}{2}\epsilon^{\mu\nu\rho} (\tilde\varepsilon^\dagger\gamma_\rho
\tilde\psi_A Y^A)\hat F_{\mu\nu}\right),
\end{align}
where $\tilde Y^A$ and $\tilde\psi_A$ were defined in (\ref{dfntil}) and
\begin{align}\label{d1Z}
\delta_1 Z_A^\dagger &= i \tilde\varepsilon^\dagger\tilde\psi_A,
\nn \\
\delta_1\psi^{\dagger A} &= D_\mu\tilde Z^A\tilde\varepsilon^\dagger
\gamma^\mu.
\end{align}
In the last step of (\ref{d1L-2}), we used the properties of SU(2)
invariant tensors, $\epsilon^{ab}$ and $\epsilon^{\hat a\hat b}$
with gauge group indices $a,b =1,2$, for instance,
\begin{align}\nn
D_\mu\tilde Z^{A\hat a}_{~~~a}
&= \partial_\mu\tilde Z^{\hat a}_{~~a} + i \hat A^{\hat a}_{\mu\hat b}
\tilde Z^{A\hat b}_{~~~a} - i \tilde Z^{A\hat a}_{~~~ b}
A^b_{\mu a}
\nn \\
&= \epsilon^{\hat a\hat b}\epsilon_{ab}\left(
\partial_\mu Z^{A b}_{~~~\hat b} + i A^b_{\mu c} Z^{Ac}_{~~~\hat b}
- i Z^{Ab}_{~~~\hat c}\hat A^{\hat c}_{\mu\hat b}
\right) \nn \\
&= \epsilon^{\hat a\hat b}\epsilon_{ab} (D_\mu Z^A)^{ b}_{~\hat b}
\end{align}
with the help of symmetric properties,
\begin{align}\nn
(\epsilon A_\mu)_{ac} &= \epsilon_{ab} A^b_{\mu c}
= \epsilon_{cb}A^b_{\mu a} = (\epsilon A_\mu)_{ca},
\nn \\
(\hat A_\mu\epsilon)^{\hat a\hat c} &= \hat A^{\hat a}_{\mu\hat b}
\epsilon^{\hat b\hat c} = \hat A^{\hat c}_{\mu\hat b}\epsilon^{\hat
b\hat a} =(\hat A_\mu\epsilon)^{\hat c\hat a}.
\end{align}

By taking supersymmetry variations for the gauge fields,
$A_\mu$ and $\hat A_\mu$, as
\begin{align}\label{dAm2}
\delta A_\mu &= \frac{2\pi}{k} \tilde\varepsilon^\dagger Z^A \gamma_\mu
\tilde\psi_A,
\nn \\
\delta\hat A_\mu &= \frac{2\pi}{k}\tilde\varepsilon^\dagger \gamma_\mu
\tilde\psi_A Z^A,
\end{align}
we obtain the same relation given in (\ref{dL1}), i.e.,
\begin{align}\label{dL1-2}
\delta_1{\cal L}_0 + \delta_A {\cal L}_{{\rm CS}}=0.
\end{align}
Now we consider the variations of the gauge fields in ${\cal L}_0$,
and obtain the following relation
\begin{align}\label{dL2-2}
\delta_A {\cal L}_0 &= {\rm tr}\Big(-i \delta_2\psi^{\dagger A}
\gamma^\mu D_\mu\psi_A\Big) +\delta_1 V_{{\rm ferm}}
\nn \\
&= - \delta_2 {\cal L}_0 + \delta_1 V_{{\rm ferm}},
\end{align}
where $V_{{\rm ferm}}$ was given in (\ref{Vferm}) and
\begin{align}\label{d2psi-2}
\delta_2\psi^{\dagger A} = - \tilde\varepsilon^\dagger\tilde N^A
\end{align}
with the definition of $\tilde A^A$ in (\ref{dfntil}).

From the variation $\delta_2\psi^{\dagger A}$ in $V_{{\rm ferm}}$,
we can obtain the bosonic potential $V_{{\rm bos}}$,
\begin{align}\label{dL3-2}
\delta_2 V_{{\rm ferm}} = -\delta_1 V_{{\rm bos}} =
-\delta V_{{\rm bos}},
\end{align}
where the expression of $V_{{\rm bos}}$ was given in (\ref{Vbos}).
During the calculational procedures in obtaining (\ref{dL2-2}) and
(\ref{dL3-2}), we frequently used the following relations
\begin{align}\nn
&\epsilon_{ab} P^a Q^b R^c = \epsilon_{ab}(-P^c Q^a + P^a Q^c)R^b,
\nn \\
&(\tilde\varepsilon^\dagger\psi_1)(\psi_2\psi_3)
=-(\tilde\varepsilon^\dagger\psi_2)(\psi_3\psi_1)
- (\tilde\varepsilon^\dagger
\psi_3)(\psi_2\psi_1),
\nn \\
&(\tilde\varepsilon^\dagger\gamma_\mu\psi_1)(\psi_2\gamma^\mu\psi_3)
= -2(\tilde\varepsilon^\dagger\psi_3)(\psi_1\psi_2) -
(\tilde\varepsilon^\dagger\psi_1) (\psi_2\psi_3),
\end{align}
where $P$, $Q$, and $R$ are arbitrary fields with gauge indices and
$\psi_1$, $\psi_2$, and $\psi_3$ are arbitrary fermion fields.

Combining the relations (\ref{dL1-2}), (\ref{dL2-2}), and (\ref{dL3-2}),
as we did in the previous subsection, we can verify the supersymmetry
transformation rules given in (\ref{SU2-1}).

\section{Supersymmetry of ABJM Theory}\label{vAsu}

\subsection{SU(2)$\times$ SU(2) case}\label{B1}
If we replace $Z$ with $Y$ and extend the SU(2) global indices
$A,B,...=1,2$ to the SU(4) global indices $A,B,... =1,2,3,4$
in (\ref{dact}), the kinetic and Chern-Simons terms are exactly same with
those in (\ref{ABJMact}), while the fermionic and bosonic potentials have
different forms in the two actions.
In this reason, in order to verify the additional ${\cal N}=2$ supersymmetry
invariance given in (\ref{U2-2}) (without $\check T$) in ABJM theory with
SU(2)$\times$SU(2) gauge group, we use the results given
in the subsection A.2. The results in the subsection A.2
are summarized in ABJM theory side, as follows,
\begin{align}\label{SUSU1}
&\delta_1{\cal L}_0 + \delta_A {\cal L}_{{\rm CS}} = 0,
\nn \\
&\delta_A {\cal L}_0 + \delta_2 {\cal L}_0 - \delta_1
V_{{\rm ferm1}} = 0,
\nn \\
&\delta_2 V_{{\rm ferm1}} + \delta_1 V_{{\rm bos1}}=0,
\end{align}
where
\begin{align}
\delta_1 Y^{\dagger}_A &= i\tilde\varepsilon^\dagger\tilde\psi_A,
\nn \\
\delta_1\psi^{\dagger A} &= D_\mu\tilde Y^A
\tilde\varepsilon^\dagger\gamma^\mu,
\nn \\
\delta_2\psi^{\dagger A} &= -\tilde\varepsilon^\dagger \tilde N^A,
\nn \\
\delta A_\mu &= \frac{2\pi}{k}\tilde\varepsilon^\dagger Y^A\gamma_\mu
\tilde\psi_A,
\nn \\
\delta \hat A_\mu &= \frac{2\pi}{k}\tilde\varepsilon^\dagger\gamma_\mu
\tilde\psi_A Y^A
\label{dYpsi-2}
\end{align}
with $\tilde N^{A\hat a}_{~~~a} =\epsilon^{\hat a\hat
b}\epsilon_{ab} \frac{2\pi}{k} ( Y^BY_B^\dagger Y^A -
Y^AY_B^\dagger Y^B)^{b}_{~\hat b}$, and
\begin{align}
V_{{\rm ferm1}} &=  \frac{2\pi i}{k} {\rm tr}
\left(Y_A^\dagger Y^A \psi^{\dagger B}\psi_B
- Y^A Y_A^\dagger\psi_B\psi^{\dagger B}
+ 2Y^A Y_B^\dagger\psi_A\psi^{\dagger B} -2 Y_A^\dagger Y^B\psi^{\dagger A}
\psi_B\right),
\label{Vferm1} \\
V_{{\rm bos1}} &= \frac{4\pi^2}{k^2} {\rm tr}\left( Y_A^\dagger Y^A
Y_B^\dagger Y^B Y_C^\dagger Y^C +Y^AY_A^\dagger Y^B Y_B^\dagger
Y^CY_C^\dagger - 2Y^AY_B^\dagger Y^BY_A^\dagger
Y^CY_C^\dagger\right).\label{Vbos1}
\end{align}

In the next, we consider a supersymmetry variation of $V_{{\rm ferm}}$
defined as
\begin{align}\label{Vferm2}
V_{{\rm ferm2}} &= V_{{\rm ferm}} - V_{{\rm ferm1}}
\nn \\
&=\frac{2\pi i}{k} {\rm tr}\left(
\epsilon^{ABCD} Y_A^\dagger\psi_B Y_C^\dagger\psi_D -
\epsilon_{ABCD} Y^A\psi^{\dagger B}Y^C\psi^{\dagger D}\right),
\end{align}
where $V_{{\rm ferm}}$ and $V_{{\rm ferm1}}$ were defined in
(\ref{ferV}) and (\ref{Vferm1}) respectively.

Then we find
\begin{align}\label{SUSU2}
\delta_1 V_{{\rm ferm2}} &= \frac{4\pi i}{k} {\rm tr}
\left(\epsilon^{ABCD} \delta_1Y_A^\dagger\psi_B Y_C^\dagger\psi_D -
\epsilon_{ABCD}\delta_1\psi^{\dagger A}Y^B\psi^{\dagger C} Y^D\right)
\nn \\
&= i \delta_3\psi_A\gamma^\mu D_\mu\psi^{\dagger A}
\nn \\
&= \partial_\mu (i \delta_3\psi_A\gamma^\mu\psi^{\dagger A}) + i
\psi^{\dagger A}\gamma^\mu D_\mu \delta_3\psi_A,
\end{align}
where
\begin{align}\label{d3psi}
\delta_3\psi_A = -\frac{4\pi}{3k}\epsilon_{ABCD}\tilde\varepsilon^\dagger
Y^B\tilde Y^C Y^D.
\end{align}
Here we used the facts that
\begin{align}\nn
&{\rm tr} \left(\epsilon^{ABCD}\delta_1 Y_A^\dagger\psi_B
Y_C^\dagger\psi_D\right)=0,
\nn \\
&{\rm tr}\left(\epsilon_{ABCD} D_\mu\tilde Y^A\tilde\varepsilon^\dagger
\gamma^\mu Y^B\psi^{\dagger C}Y^D\right)=
-\frac{1}{3} {\rm tr} \left(\epsilon_{ABCD}Y^B\tilde Y^C Y^D
\tilde\varepsilon^\dagger\gamma^\mu D_\mu \psi^{\dagger A}\right).
\end{align}
After slightly long algebra, we obtain
\begin{align}\label{SUSU3}
\delta_2 V_{{\rm ferm2}} = -\delta_3 V_{{\rm ferm1}},
\end{align}
where
\begin{align}\nn
\delta_2 V_{{\rm ferm2}} &= \frac{4\pi i}{k} {\rm tr}
\left(\epsilon_{ABCD}\delta_2\psi^{\dagger A} Y^B\psi^{\dagger C} Y^D\right),
\nn \\
\delta_3 V_{{\rm ferm1}} &= \frac{2\pi i}{k} {\rm tr}
\left(\delta_3\psi_A Y_B^\dagger Y^B\psi^{\dagger A} - \delta_3\psi_A
\psi^{\dagger A}Y^B Y_B^\dagger + 2\delta_3\psi_A
\psi^{\dagger B} Y^A Y_B^\dagger - 2\delta_3\psi_A Y_B^\dagger Y^A
\psi^{\dagger B}\right).\nn
\end{align}

As a final step we check $\delta_3 V_{{\rm ferm2}}$ and find the
corresponding bosonic potential.
By expending $\epsilon^{ABCD}\epsilon_{AEFG}$ and
using the properties of the SU(2) invariant tensors, we finally find the
following relation
\begin{align}\label{SUSU4}
\delta_3 V_{{\rm ferm2}} = - \delta_1 V_{{\rm bos2}},
\end{align}
where
\begin{align}\nn
\delta_3 V_{{\rm ferm2}} &= -\frac{4\pi i}{k} {\rm tr}
\left(\epsilon^{ABCD}\delta_3\psi_A Y_B^\dagger\psi_C Y_D^\dagger\right),
\nn \\
V_{{\rm bos2}} &= -\frac{16\pi^2}{3k^2}{\rm tr}\left(
Y_A^\dagger Y^A Y_B^\dagger Y^B Y_C^\dagger Y^C + Y^A Y_A^\dagger
Y^BY_B^\dagger Y^CY_C^\dagger
\right. \nn \\
&\left.\hskip 2.1cm + Y_A^\dagger Y^B Y_C^\dagger
Y^AY_B^\dagger Y^C -3 Y^A Y_B^\dagger Y^B Y_A^\dagger Y^C Y_C^\dagger\right).
\end{align}
From these results we find
\begin{align}\nn
V_{{\rm bos1}} +V_{{\rm bos2}} =V_{{\rm bos}},
\end{align}
where $V_{{\rm bos}}$ is the bosonic potential of ABJM theory.
From (\ref{SUSU1}), (\ref{SUSU2}), (\ref{SUSU3}), and (\ref{SUSU4}),
we prove that the ABJM action (\ref{ABJMact}) is invariant
under the supersymmetry transformation rule (\ref{U2-2})
(without $\check T$), which can be obtained from (\ref{dYpsi-2})
and (\ref{d3psi}), and their complex conjugates.

\end{document}